\documentclass[aps,prl,twocolumn,superscriptaddress]{revtex4-2}

\usepackage[utf8]{inputenc}   
\usepackage[T1]{fontenc}      
\usepackage[english]{babel}

\usepackage{amsmath,amssymb}
\usepackage{graphicx}
\usepackage[colorlinks=true,linkcolor=blue,citecolor=blue,urlcolor=blue]{hyperref}
\usepackage{amsmath,amssymb,hyperref,graphicx,color}
\usepackage{caption}
\usepackage{subcaption}
\usepackage{braket}
\usepackage{ragged2e}

\definecolor{rossocorsa}{rgb}{0.83, 0.0, 0.0}
\definecolor{bleudefrance}{rgb}{0.19, 0.55, 0.91}
\usepackage{hyperref}
\hypersetup{
    colorlinks,
    citecolor=bleudefrance,
    filecolor=bleudefrance,
    linkcolor=rossocorsa,
    urlcolor=bleudefrance
}
\usepackage[normalem]{ulem}

\newcommand{\stkout}[1]{\ifmmode\text{\sout{\ensuremath{#1}}}\else\sout{#1}\fi}
\DeclareMathOperator{\sech}{sech}
\newcommand{\be}{\begin{equation}}
\newcommand{\ee}{\end{equation}}
\newcommand{\bea}{\begin{eqnarray}}
\newcommand{\eea}{\end{eqnarray}}
\newcommand{\beas}{\begin{eqnarray*}}
\newcommand{\eeas}{\end{eqnarray*}}

\newcommand{\suniv}{\mathsf S_{\text{u}}}

\newcommand{\Ls}{L_\star}
\newcommand{\diff}{\text{d}}

\newcommand{\tE}{t_\text{E}}
\newcommand{\iu}{\text i}
\newcommand{\tauE}{\tau_\text{E}}

\newcommand{\Tau}{\mathrm{T}}

\newcommand{\TauEl}{\mathrm{T}_{\text E,\ell}}

\begin{document}

\title{Universality of pseudoentropy for deformed spheres in dS/CFT}

\author{Giorgos Anastasiou}
\affiliation{Departamento de Ciencias, Facultad de Artes Liberales, Universidad Adolfo Ibáñez, \\ Avenida Diagonal Las Torres 2640, 7941169, Pe\~nalol\'en, Chile}
\author{Ignacio J. Araya}
\affiliation{Departamento de Física y Astronomía, Facultad de Ciencias Exactas, Universidad Andres Bello, Sazi\'e 2212, Piso 7, Santiago, Chile}
\author{Avijit Das}
\affiliation{Instituto de Ciencias Exactas y Naturales, Universidad Arturo Prat, Avenida Playa Brava 3256, 1111346, Iquique, Chile}
\affiliation{Facultad de Ciencias, Universidad Arturo Prat, Avenida Arturo Prat Chacón 2120, 1110939, Iquique, Chile}
\author{Javier Moreno}
\affiliation{Center for Gravitational Physics and Quantum Information, Yukawa Institute for Theoretical Physics, Kyoto University, Kitashirakawa Oiwakecho, Sakyo-ku, Kyoto 606-8502, Japan.}
\affiliation{Departament de Física Quàntica i Astrofísica, Institut de Ciències del Cosmos, Universitat de Barcelona, Martí i Franquès 1, E-08028 Barcelona, Spain.}

\begin{abstract}
\noindent We determine the universal part of pseudoentropy for small shape deformations of spherical entangling surfaces in the context of de Sitter/conformal field theory (dS/CFT) correspondence. The leading correction at quadratic order in the deformation parameter is controlled by the analytic continuation of the coefficient of the two-point stress-energy tensor correlator in AdS/CFT (i.e., $\left.\Ls\right|_{\text{AdS}}\rightarrow-\iu \left.\Ls\right|_{\text{dS}}$), thereby establishing the sphere as a local extremum. The same structure holds in higher-curvature theories, as we check explicitly for quadratic curvature gravity, suggesting a universal behavior across non-unitary holographic CFTs. Our findings extend the Mezei formula to the dS/CFT setting and indicate that the shape dependence of pseudoentropy in dS holography resembles that of entanglement entropy in AdS space. Thus, we conjecture this coefficient to be the $C_T$ for the non-unitary CFT dual.
\end{abstract}

\maketitle

\noindent\emph{Introduction}. Entanglement entropy, defined as the von Neumann entropy of a reduced density matrix in a bipartite Hilbert space, provides a fundamental measure of quantum correlations and plays a central role in quantum information theory and quantum gravity---see~\cite{Bombelli:1986rw,Srednicki:1993im,Calabrese:2004eu,Ryu:2006bv,VanRaamsdonk:2010pw,Lewkowycz:2013nqa}, among others. To extend this notion to pairs of non-identical states, pseudoentropy was introduced in~\cite{Nakata:2020luh}, defined through the reduced transition matrix as
\begin{equation}
S(A)=-\text{tr}(\tau_A\log\tau_A)\,,\quad\tau_A=\text{tr}_B\left(\frac{|\psi\rangle\langle\varphi|}{\braket{\varphi|\psi}}\right)\,,
\end{equation}
where $|\psi\rangle$ and $|\varphi\rangle$ are pure states in the total Hilbert space $\mathcal{H}_{\text{tot}}=\mathcal{H}_{A}\otimes \mathcal{H}_{B}$, and $\tau_A$ is the reduced transition matrix obtained by tracing out subsystem $B$. The resulting quantity $S(A)$ is the pseudoentropy of subsystem $A$. Here $\tau_A$ is generally non-Hermitian \cite{Doi:2022iyj,Doi:2023zaf}, so $S(A)$ can take complex values. This property makes pseudoentropy a useful information measure for non-Hermitian quantum systems in condensed matter physics \cite{Couvreur:2016mbr,Herviou:2019yfb,Chang:2019jcj,Jian:2020byi}.

In the holographic framework, pseudoentropy emerges as a prominent quantity within the de Sitter/conformal field theory (dS/CFT) correspondence \cite{Strominger:2001pn,Witten:2001kn,Maldacena:2002vr}, where gravity in a dS spacetime is conjectured to be dual to a Euclidean CFT defined on its boundary at future time infinity. Unlike its anti-de Sitter counterpart (AdS/CFT) \cite{Maldacena:1997re,Gubser:1998bc,Witten:1998qj}, the Euclidean CFT dual to dS space is non-unitary \cite{Maldacena:2002vr}, and only a limited number of explicit realizations of this duality have been constructed so far \cite{Anninos:2011ui,Mollabashi:2021xsd,Hikida:2022ltr}. In this setting, the holographic entanglement entropy \cite{Ryu:2006bv,Ryu:2006ef,Nishioka:2009un} generally acquires complex values and is thus interpreted as a pseudoentropy \footnote{Additionally, the recently proposed timelike entanglement entropy, defined for timelike-separated regions in Lorentzian field theories, can also take complex values and is naturally interpreted as a pseudoentropy \cite{Liu:2022ugc,Doi:2022iyj,Doi:2023zaf,Narayan:2022afv,Narayan:2023ebn,Das:2023yyl,Heller:2024whi}}---see also~\cite{Sanches:2016sxy,Arias:2019pzy,Ruan:2025uhl}.

\begin{figure}
    \centering
    \includegraphics[width=0.8\linewidth]{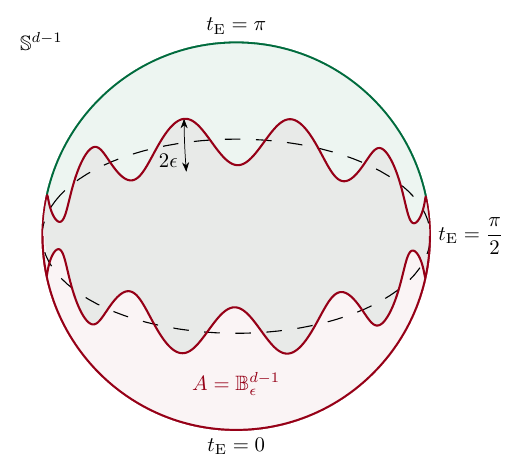}
    \caption{\justifying In green, the entangling region $\mathbb B^{d-1}_\epsilon$ is the perturbation around the   unit-ball $\mathbb B^{d-1}$, stereographically projected on $\mathbb S^{d-1}$.}
    \label{fig:setup}
\end{figure}

In the dS/CFT correspondence, the partition function of the non-unitary Euclidean CFT is dual to the Hartle-Hawking wavefunction in dS space \cite{Maldacena:2002vr}. Starting with global dS$_{d+1}$ coordinates, the metric $\diff s^2=g_{\mu\nu}\diff x^\mu\diff x^\nu$ reads
\begin{equation}\label{eq:GdS1} 
    \diff s^2=\Ls^2\left[-\diff\tau^2+\cosh^2\tau\left(\diff\tE^2+\sin^2\tE\diff\Omega_{d-1}^2\right)\right]\,,
\end{equation}
where $\Ls$ is the dS radius, $\tau$ is the bulk time coordinate, $\tE$ is the Euclidean time of the CFT \footnote{The compact Euclidean direction $\tE \in[0, \pi]$ serves as one of the angular coordinates on the global boundary 
$\mathbb S^{d}$. It arises from performing  Wick rotation on the Lorentzian time $t$ in the AdS formulation.} and $\diff\Omega_{d-1}^2=\diff\theta^2+\cos\theta^2\,\diff\Omega_{d-2}^2$ is the metric of the unit $\mathbb S^{d-1}$. The Hartle-Hawking initial state is then prepared by the path integral on half of the Euclidean dS$_{d+1}$, obtained via  Wick-rotating the bulk time $\tau=-\iu\tauE$. Consequently, in the range $\tauE\in[0,\pi/2]$ the geometry is
\begin{equation}\label{eq:GdT1} 
    \diff s^2=\Ls^2\left[\diff\tauE^2+\cos^2\tauE\left(\diff\tE^2+\sin^2\tE\diff\Omega_{d-1}^2\right)\right]\,,
\end{equation}
and the full geometry is properly interpreted as a gluing of Lorentzian dS$_{d+1}$ and Euclidean dS$_{d+1}$, joined smoothly at the hypersurface $\tau=\tauE=0$ \footnote{The mixed nature of the bulk manifold is a consequence of the Hartle-Hawking geometry associated with the `no-boundary proposal' in dS/CFT holography\cite{Strominger:2001pn,Maldacena:2002vr}}.

In this setup, the pseudoentropy associated with a subregion $A$ on the boundary $\mathbb S^d$ \cite{Doi:2022iyj,Doi:2023zaf}, can be computed holographically by extending the Ryu-Takayanagi (RT) prescription to dS space---see Eq.~\eqref{eq:RTformula} below---which leads to a complex-valued codimension-two area functional in the bulk. Moreover, in \cite{Doi:2022iyj,Doi:2023zaf}, it was shown that for a ball-shaped entangling region $A=\mathbb B^{d-1}$ with entangling surface $\partial A=\mathbb S^{d-2}$, defined on the $\theta=0$ slice, the universal part of holographic pseudoentropy, i.e., 
\begin{equation}\label{eq:Sexp}
    S(\mathbb B^{d-1})\supset\begin{cases}
     (-)^{\frac{d-1}{2}}\suniv(\mathbb B^{d-1}) & \text{if odd }d\,,\\
      (-)^{\frac{d-2}{2}}\, \suniv(\mathbb B^{d-1}) (\log\frac{T_0}{\delta} -\frac{\iu \pi}{2})& \text{if even }d\,,
   \end{cases}
\end{equation}
where $T_0$ is a characteristic length of the subregion and $\delta$ is an ultraviolet regulator, contains information regarding the complex-valued central charges of the dual Euclidean CFT$_d$. Notice that in even dimensions there is a universal contribution at finite order, which is not contaminated by a physical redefinition of $\delta$. This term relates to timelike entanglement entropy by analytic continuation of $T_{0}$.

Moving away from the ball-shaped region, we consider a slightly deformed ball $A = \mathbb B^{d-1}_\epsilon$ of unit radius, where $\epsilon \ll 1$ is a small deformation parameter. The shape deformation is represented by the profile
\begin{equation}\label{eq:def}
    \mathbb B_\epsilon^{d-1}:\ T_0=\frac{\pi}{2} + \epsilon \sum_{\ell,\mathbf m} a_{\ell,\mathbf m}Y_{\ell,\mathbf m}\,, 
\end{equation}
where $Y_{\ell,\mathbf m}=Y_{\ell,\mathbf m}(\Omega_{d-2})$ is the $(d-2)$-spherical harmonic with quantum numbers $\ell$, $\mathbf{m}=\{m_1,\ldots m_{d-3}\}$, satisfying $\Delta Y_{\ell,\mathbf m}=-\ell(\ell+d-3)Y_{\ell,\mathbf m}$, with $\Delta$ denoting the Laplace-Beltrami operator on $\mathbb S^{d-2}$, and $a_{\ell,\mathbf m}$ the amplitude of the deformation. In general, the Euclidean time coordinate ranges as $\tE\in[0,\pi]$. A fixed value of $\tE$ specifies a ball-shaped entangling subregion of radius $r=\tan\left(\tE/2\right)$ via stereographic projection. In particular, the value $\tE=T_0=\pi/2$ corresponds to the unit-radius case of interest. Figure~\ref{fig:setup} illustrates the deformation profile \eqref{eq:def}.

Using this setup, in this Letter, we conjecture that, for a broad class of non-unitary CFTs---namely, those admitting a dual description in terms of some gravitational theory---the universal part of the pseudoentropy admits the expansion
\begin{equation}\label{eq:Sdef1}
\suniv(\mathbb B_\epsilon^{d-1})=\suniv^{(1)}+\epsilon^2\suniv^{(2)}\,,\ \ \suniv^{(1)}=\begin{cases}
    2\pi a^\star & \text{if odd }d\,,\\
    4a^\star & \text{if even }d\,,
\end{cases}
\end{equation}
where $a^\star=(-\iu)^{d-1}\pi^{(d-2)/2}\Ls^{d-1}/[8\Gamma(d/2)G]$ is the complex-valued central charge of the non-unitary CFT$_d$ \cite{Maldacena:2002vr} with the leading correction, at quadratic order in $\epsilon$, reading
\begin{widetext}
\begin{equation}\label{eq:Sdef2}
\suniv^{(2)}=C_T\frac{\pi^{(d+2)/2}(d-1)}{2^{d-2}\Gamma(d+2)\Gamma(d/2)}\sum_{\ell,\mathbf m} a_{\ell,\mathbf m}^2(\ell-1)_d\times\begin{cases}
    \pi/2& \text{if odd }d\,,\\
    1 & \text{if even }d\,,
\end{cases}
\end{equation}  
\end{widetext}
where $(x)_n=\prod_{k=1}^n(x+k-1)$ is the Pochhammer symbol. The coefficient $C_T$---see Eq.~\eqref{eq:CT} below---characterizes the two-point stress energy tensor correlator \cite{Osborn:1993cr}, $\langle T_{\alpha\beta}(x)T_{\gamma\delta}(0)\rangle=C_T/x^{2d}\left(I_{\alpha(\gamma}I_{\delta)\beta}-\delta_{\alpha\beta}\delta_{\gamma\delta}/d\right)$, where $I_{\alpha\beta}=\delta_{\alpha\beta}-2x_\alpha x_\beta/x^2$ \footnote{Throughout this Letter we use: i) $\alpha$, $\beta$, $\gamma$, $\delta$ for indices in the conformal field theory, ii) $\mu,\nu,\rho,\sigma$ as indices for the dual gravity theory, iii) $a$, $b$ for the codimension-two surface $\Sigma_A$ and $i$, $j$ for normal directions to $\Sigma_A$.}. As opposed to unitary CFTs, where $C_T>0$, this may vanish, become negative, or even take complex values in non-unitary theories---see e.g., \cite{Osborn:2016bev,Flohr:2001zs,Buchel:2009sk,Myers:2010jv,Anninos:2011ui}.

Our main result, Eq.~\eqref{eq:Sdef2}---together with Eq.~\eqref{eq:CT} and Eq.~\eqref{eq:CTQCG} below, extends Mezei’s formula for entanglement entropy  \cite{Allais:2014ata,Mezei:2014zla} to a universality class of non-unitary CFTs with gravity duals, showing that the pseudoentropy of a perturbatively deformed spherical entangling surface still satisfies the same structural relation as in the unitary AdS/CFT case.

\noindent\emph{Holographic pseudoentropy for perturbed spheres}. For holographic Euclidean CFTs dual to Einstein-dS defined at future time infinity, the pseudoentropy is given by the RT formula \cite{Ryu:2006bv,Ryu:2006ef}
\begin{equation}\label{eq:RTformula}
    S(A)=\frac{\text{Area}(\Sigma_A)}{4G}\,,
\end{equation}
where the codimension-two surface $\Sigma_A$ is extremal in the dS bulk. The pseudoentropy area functional in dS space contains both timelike and spacelike segments, i.e., $\Sigma_A=\Sigma_A^{(\text t)}\cup\Sigma_A^{(s)}$, and consequently $S(A)=S^{(\text t)}(A)+S^{( \text s)}(A)$ \footnote{The piecewise construction is not generically valid in higher dimensions or for arbitrary subsystems; in particular, it can fail for strip-like regions even in AdS \cite{Doi:2023zaf}. However, for spherical subregions, it provides a consistent representative of the relevant saddle in pure dS ~\cite{Doi:2023zaf,Narayan:2026wzp}. We posit that this is still the case for small deformations thereof.}. The timelike part, $\Sigma_A^{(\text t)}$, is embedded in the Lorentzian section of the bulk dS$_{d+1}$ geometry \eqref{eq:GdS1}. On the other hand, the spacelike part $\Sigma_A^{(\text s)}$, is embedded in the Euclidean section of the bulk dS$_{d+1}$ given in Eq.~\eqref{eq:GdT1}.

The extremal surface $\Sigma_A$ is determined by requiring the trace of the extrinsic curvature along each normal direction $n_\mu^i$ to vanish separately. Explicitly, $K^{i}=h^{ab}K^{i}{}_{ab}=h^{ab}h_a^\mu h_b^\nu\nabla_\nu n_\mu^{i}$, where $h_{ab}$ is the induced metric, $h_a^\mu$ its projector and $i$ labels the independent spacelike and timelike normals to the codimension-two surface $\Sigma_A$. Using the embedding function $\tE=\tE(\tau,\Omega_{d-2})$ for the timelike part, and, $\tE=\tE(\tauE,\Omega_{d-2})$ for the spacelike part, one can separate variables at linear order in the perturbation parameter as
\begin{equation}\label{eq:tE}
    \tE=\frac{\pi}{2}+\epsilon\sum_{\ell,\mathbf m}a_{\ell,\mathbf m}Y_{\ell,\mathbf m}\times\begin{cases}
        \Tau_{\ell}(\tau) & \text{for } \tau\in(\infty,0)\,,\\
        \TauEl(\tauE) & \text{for } \tauE\in(0,\frac{\pi}{2})\,.
    \end{cases}
\end{equation}
Then, the vanishing of the trace of the extrinsic curvature associated to the normal of the $\tE(\tau,\Omega_{d-2})$ surface leads to the differential equation for the timelike part \footnote{Notice that the extrinsic curvature along the $\theta$ direction vanishes identically, $  K^{\theta}{}_{ab}  = h^\mu_a h^\theta_b \nabla_\mu n_{\theta}^{\theta} =0$ since, for any value of $b$, $ h^\theta _b = 0$.},

\begin{equation}\label{eq:diffeq1}
\left[c_\ell\sech^2{\tau} + d \tanh{\tau}\frac{\diff}{\diff \tau}+ \frac{\diff^2}{d \tau^2}\right]\Tau_{\ell}(\tau)=0\,,
\end{equation}
where $c_\ell=\ell^2 + (d - 3) \ell - (d - 2)$. On the other hand, for the spacelike part, it suffices to find the Wick-rotated version of the timelike one, obtaining
\begin{equation}\label{eq:diffeq2}
\left[c_{\ell} \sec^2{\tauE} + d \tan{\tauE}\frac{\diff}{\diff \tauE}- \frac{\diff^2}{d \tauE^2}\right]\TauEl(\tauE)=0\,.
\end{equation}

\begin{figure*}
\centering
\begin{subfigure}{.5\textwidth}
  \centering
  \includegraphics[height=0.6\linewidth]{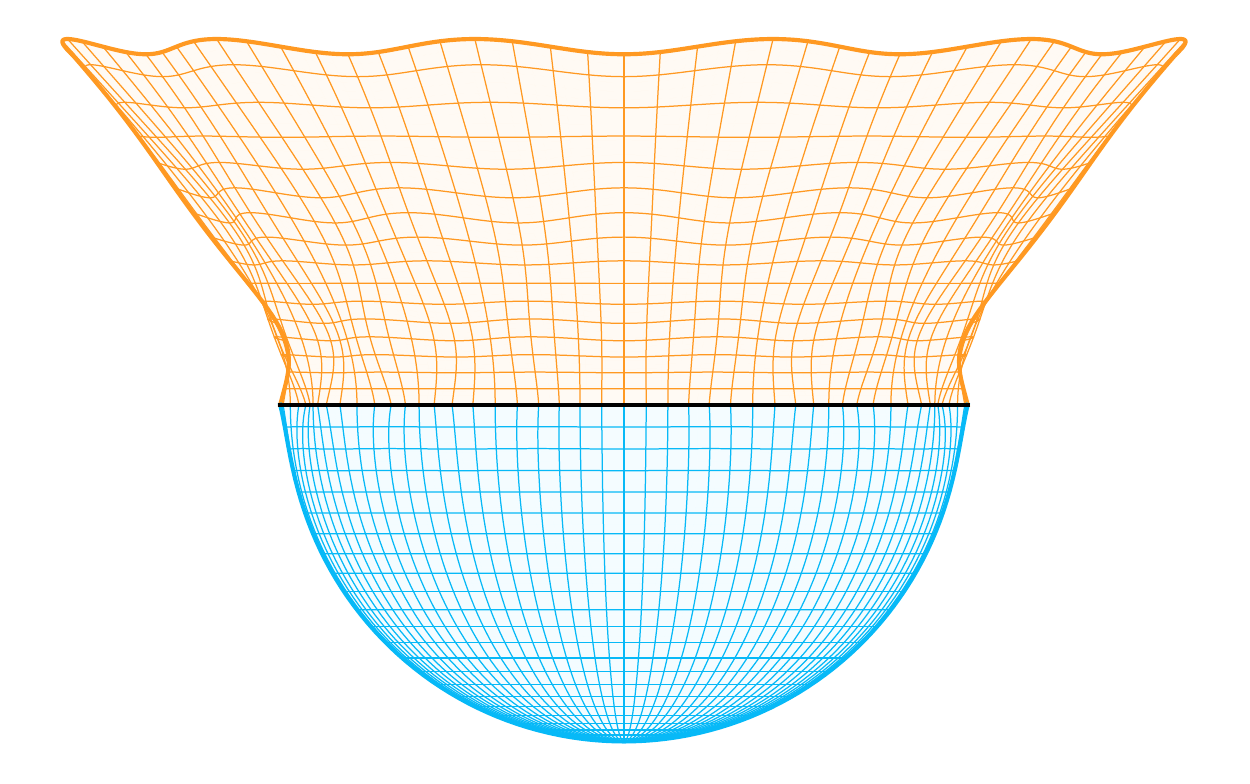}
\end{subfigure}%
\begin{subfigure}{.5\textwidth}
  \centering
  \includegraphics[height=0.6\linewidth]{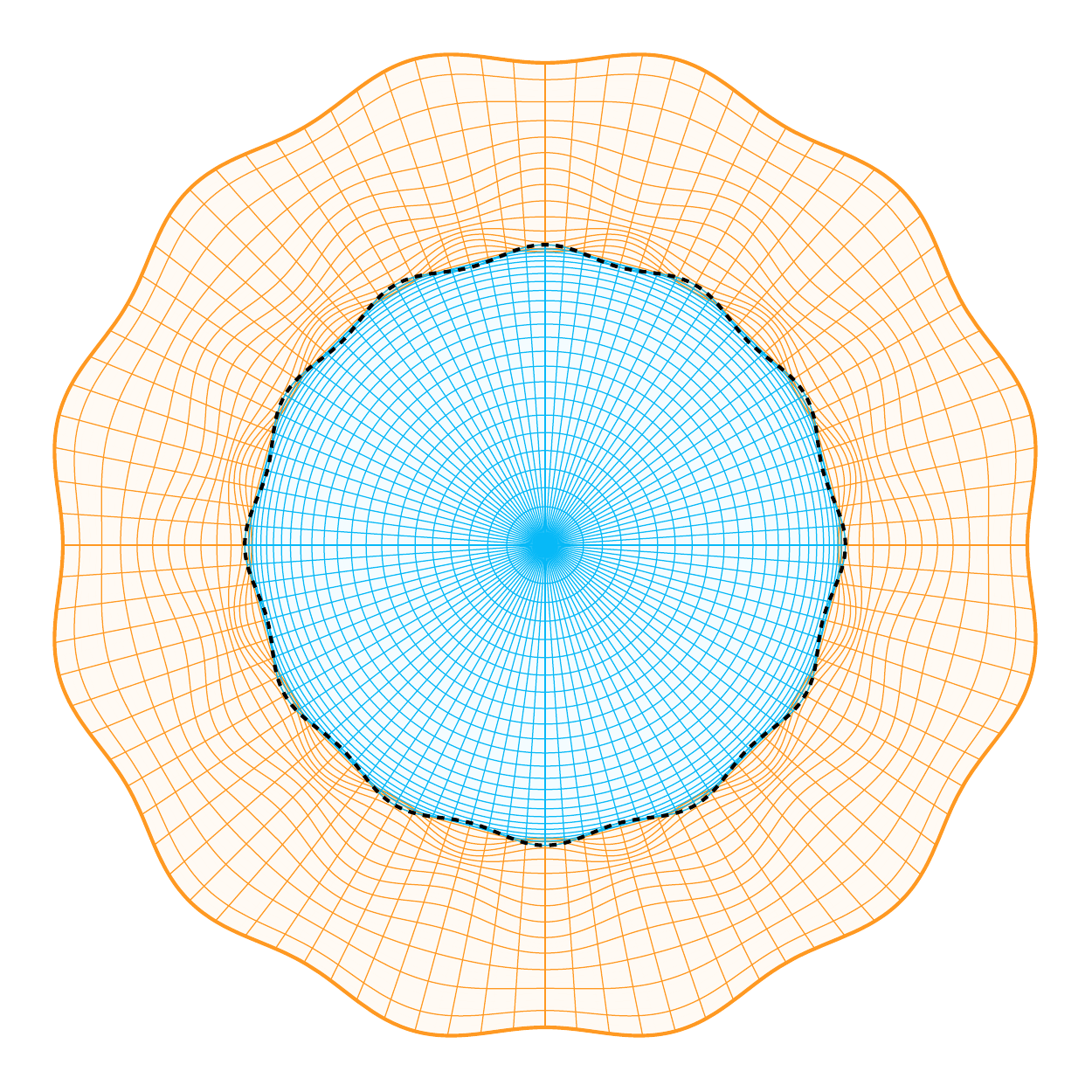}
  \label{fig:sub2}
\end{subfigure}
\caption{\justifying(Left) Front view of the RT surface for small perturbations around the ball-shaped entangling region at future infinity. (Right) Top view of the same RT surface. The timelike section is shown in orange, while the spacelike part is highlighted in blue. The black line indicates the junction at $\tau = \tau_E = 0$. The sketch is not conformal; therefore, the upper part does not correspond to future infinity, but to some value of $0<\tau<\infty$.}
\label{fig:test}
\end{figure*}

The (four) undetermined coefficients from these second order differential equations are obtained using: i) the boundary conditions $\Tau_\ell(\tau\rightarrow\infty)=1$ and $\TauEl(\tauE=\pi/2)=0$, which correspond to fixing the radius at the conformal boundary and the turning point, respectively, and, ii) the junction conditions at $\tau = \tau_E = 0$, namely $\Tau_\ell(0)=\TauEl(0)$ and $\partial_{\tau}\Tau_\ell(0)=\iu\partial_{\tauE}\TauEl(0)$ \footnote{In the case of the undeformed sphere, the second junction (regularity) condition is unnecessary, since the final pseudoentropy is independent of the coefficient that would otherwise remain undetermined by this additional requirement \cite{Doi:2023zaf}.}. These conditions ensure continuity and differentiability as we move 
from the timelike to the spacelike segments of the RT surface. The smoothness requirement arises due to the absence of localized matter at $\tau=\tauE=0$, in analogy to the usual junction conditions of \cite{Israel:1966rt}. Imposing the conditions we obtain for the timelike part
\begin{align}\label{eq:tauT} 
    \Tau_\ell(\tau) =&\frac{(-)^{3d/2}\sech^{d/2}\tau}{2^{d/2}\Gamma(d/2)} \left[ \iu\pi P_{\ell+d/2-2}^{d/2}\left(\tanh\tau\right)\right. \notag\\
    &+ \left.2Q_{\ell+d/2-2}^{d/2}\left(\tanh\tau\right)\right]\,,
\end{align}
where $P_n^m(x)$ and $Q_n^m(x)$ are the associate Legendre polynomials of the first and second kind, respectively. On the other hand, the spacelike profile reads 
\begin{align}\label{eq:tauS} 
&\TauEl(\tauE) =\frac{(-)^{(\ell-1)/2}\sqrt{\pi}\cos^{\ell-1}\tauE}{\Gamma(d/2)}\times\\
&\left[\frac{\Gamma[(d+\ell-1)/2]}{\Gamma(\ell/2)}\right.{}_2F_1\left(\frac{\ell-1}{2},\frac{d+\ell-1}{2};\frac{1}{2};\sin^2\tauE\right)\notag\\
&\left.-\frac{2\Gamma[(d+\ell)/2]}{\Gamma[(\ell-1)/2]}{}_2F_1\left(\frac{\ell}{2},\frac{d+\ell}{2};\frac{3}{2};\sin^2\tauE\right)\sin\tauE\right]\,,\notag
\end{align}
where ${}_2F_1$ is the Gaussian hypergeometric function. In Figure~\ref{fig:test} we plot the front and top views of the perturbed RT surfaces associated to the deformed spherical entangling surface in the $d=3$ case, in which the small perturbations \eqref{eq:def} are given in a single angular coordinate \footnote{The surface takes on a geometry reminiscent of a shuttlecock, with $\ell$ corresponding to the number of feathers in its skirt.}.

Once the embedding functions for the timelike and spacelike components of the deformed RT surface are determined---namely, Eq.~\eqref{eq:tE} together with Eqs.~\eqref{eq:tauT} and \eqref{eq:tauS} for $\Tau_\ell(\tau)$ and $\TauEl(\tauE)$, respectively—they can be substituted into the RT formula~\eqref{eq:RTformula}. Starting with odd dimensions and expanding up to $\epsilon^2$, we observe that the finite contribution from $\Sigma_A^{(\text t)}$ to the universal part of the pseudoentropy becomes
\begin{equation}\label{eq:Stodd}
{\suniv^{(\text t)}}(\mathbb B_\epsilon^{d-1})
=\epsilon^2\sum_{\ell,\mathbf m}a^2_{\ell,\mathbf m}[1+(-)^{\ell}]
\frac{(-\iu)^{d-1}\pi(\ell-1)_d}{2^{d+2}\Gamma(d/2)^2}\frac{\Ls^{d-1}}{G}\,,
\end{equation}
while, for the contribution from $\Sigma_A^{(\text s)}$ to the pseudoentropy, we find
\begin{equation}\label{eq:Ssodd}
{\suniv^{(\text s)}}(\mathbb B_\epsilon^{d-1})
=2\pi a^\star-\epsilon^2\sum_{\ell,\mathbf m}a^2_{\ell,\mathbf m}
\frac{(-)^{\ell}(-\iu)^{d-1} \pi(\ell-1)_d}{2^{d+2}\Gamma(d/2)^2}\frac{\Ls^{d-1}}{G}\,.
\end{equation}
Conversely, in even dimensions, the situation is more subtle due to the presence of both the logarithmic divergence and the finite piece, each carrying universal information. In these dimensions, we find that the contributions  from $\Sigma_A^{(\text t)}$ and $\Sigma_A^{(\text s)}$, respectively read%
\begin{widetext}
\begin{align} \label{eq:Steven}
&S^{(\text t)}(\mathbb B_\epsilon^{d-1})\supset(-1)^{\frac{d-2}{2}}\left[4a^\star \log\frac{T_0}{\delta}+\left(\log\frac{T_0}{\delta}-[1+(-)^{\ell}]\frac{\iu\pi}{2}\right)\epsilon^2\sum_{\ell,\mathbf m}a^2_{\ell,\mathbf m}
\frac{(-\iu)^{d-1}(\ell-1)_d}{2^{d+1}\Gamma(d/2)^2}\frac{\Ls^{d-1}}{G}\right]\,,\\ \label{eq:Sseven}
&{S^{(\text s)}}(\mathbb B_\epsilon^{d-1})
\supset-(-1)^{\frac{d-2}{2}}\frac{\iu\pi}{2}
\left[4a^\star-\epsilon^2\sum_{\ell,\mathbf m}a^2_{\ell,\mathbf m}
\frac{(-)^{\ell}(-\iu)^{d-1}(\ell-1)_d}{2^{d+1}\Gamma(d/2)^2}\frac{\Ls^{d-1}}{G}\right]\,,
\end{align}
\end{widetext}
where we do not include nonuniversal finite terms that can be polluted by a redefinition of $\delta$. 

Interestingly, in both even and odd dimensions, we
find: i) real, finite contributions from both timelike and spacelike parts of the RT surface---in contrast with the undeformed case, where the finite piece comes only from the spacelike section, ii) parity-dependent terms with opposite signs. Thus, upon combining Eqs.~\eqref{eq:Stodd} and \eqref{eq:Ssodd} in odd dimensions, and Eqs.~\eqref{eq:Steven} and \eqref{eq:Sseven} in even dimensions, these terms cancel exactly, yielding the universal contributions given in \eqref{eq:Sdef1} and \eqref{eq:Sdef2}.  In the latter expressions, we identify a coefficient
\begin{equation}\label{eq:CT}
C_T=\frac{(-\iu)^{d-1}\Gamma(d+2)}{8\pi^{(d+2)/2}(d-1)\Gamma(d/2)}\frac{\Ls^{d-1}}{G}\,,
\end{equation}
thus obtaining an analogous expansion in dS as the Mezei formula for the entanglement entropy in AdS/CFT \cite{Allais:2014ata,Mezei:2014zla,Faulkner:2015csl}. The relationship between the dS and AdS formalisms can also be established through an analytic continuation, defined by $\left.\Ls\right|_{\text{AdS}}\rightarrow-\iu \left.\Ls\right|_{\text{dS}}$, which implies $\left.a^\star\right|_{\text{AdS}}\rightarrow\left.a^\star\right|_{\text{dS}}$. In direct analogy, we see from Eq.~\eqref{eq:CT} that the coefficient $C_T$ transforms in the same way: $\left.C_T\right|_{\text{AdS}}\rightarrow\left.C_T\right|_{\text{dS}}$. This motivates the identification of $C_T$ as
the coefficient of the two-point stress-energy tensor for CFTs dual to Einstein-dS gravity.

The result of $C_T$ obtained through analytic continuation can be motivated independently by Ref.~\cite{Ghodsi:2014hua}, where it was computed for Conformal Gravity on AdS spacetimes.  One can select Einstein-AdS or Einstein-dS from Conformal Gravity by imposing a Neumann boundary condition on the metric expansion~\cite{Maldacena:2011mk,Hell:2023rbf}. Therefore, it is reasonable to expect that the $C_T$ result for dS can be directly obtained by the analytic continuation on $\left.L_\star\right|_{\text{AdS}}$, in agreement with what is found in Eq.~\eqref{eq:CT}. 

\noindent\emph{Higher-curvature gravity}. To test the universality of Eq.~\eqref{eq:Sdef2}, we extend our analysis to non-unitary CFTs dual to higher-curvature gravities. Since the replica trick induces the same orbifold structure in both dS and AdS spacetimes \cite{Nanda:2025tid}, one should expect pseudoentropy in dS/CFT, for general gravity theories $\mathcal L=\mathcal L(g_{\mu\nu}, R_{\mu\nu\rho\sigma})$, to be captured by the Dong–Camps formula \cite{Dong:2013qoa,Camps:2013zua} and apply it to the perturbative regime \footnote{While, for generic entangling surfaces, cubic curvature invariants are affected by the splitting problem \cite{Miao:2015iba,Miao:2014nxa,Camps:2014voa}, which introduces ambiguities in the regularization of the conical singularity---and hence in the resulting holographic entropy functional, this issue does not arise at the perturbative level \cite{Dong:2013qoa,Fursaev:2013fta,Camps:2013zua,Bueno:2020uxs,Anastasiou:2021swo}.}. We hereby analyze one of the simplest examples of higher-curvature gravity, in which universality properties can be probed, namely, quadratic gravity. This theory is defined by the sum of the three quadratic contractions of the Riemann tensor on top of the Einstein-Hilbert term, i.e., $\mathcal L_{\text{QG}}=R-2\Lambda+\lambda_1 R^2+\lambda_2 R_{\mu\nu}R^{\mu\nu}+\lambda_3 R_{\mu\nu\rho\sigma}R^{\mu\nu\rho\sigma}$, where $\lambda_1$, $\lambda_2$ and $\lambda_3$ are dimensionful, undetermined coupling constants.

Using the results of \cite{Fursaev:2013fta} for scalars of quadratic order in the Riemannian curvature evaluated on squashed cones---such as the replica orbifold, the entanglement entropy functional is given by
\begin{align}
S_{\text{QG}}(A)=&\frac{1}{4G}\int_{\Sigma_A}\left[1+2\lambda_1 R + \lambda_2\left(R_i{}^i -\frac{1}{2} K_i K^i\right)\right.\notag\\
&\left.+ 2 \lambda_3 \left(R_{ij}{}^{ij} - K^i{}_{ab}K_i{}^{ab}\right)\right]\,,\label{eq:SEEQCG}
\end{align}
where $R_i{}^i=R^{\mu\nu}n_{\mu}^i n^{j}_\nu\delta_{ij}$ is the Ricci tensor projected along the normal directions and $R_{ij}{}^{ij}=R^{\mu\nu\rho\sigma}n_{\mu}^i n^{j}_\nu n_{\rho}^k n^{l}_\sigma\delta_{ik}\delta_{jl}$ is the corresponding projection of the Riemann tensor. Of course, setting $\lambda_1=\lambda_2=\lambda_3=0$ in Eq.~\eqref{eq:SEEQCG} retrieves the RT formula in Eq.~\eqref{eq:RTformula}.

To evaluate Eq.~\eqref{eq:SEEQCG}, we note that dS space is maximally symmetric, with curvature tensor $R_{\mu\nu\rho\sigma} = (g_{\mu\rho} g_{\nu\sigma} - g_{\mu\sigma} g_{\nu\rho})/\Ls^2$, where $\Ls$  now denotes the effective dS radius. Using this expression together with the Gauss–Codazzi–Mainardi relations \cite{Anastasiou:2025dex}, we substitute $R = d(d+1)/\Ls^2$, $R_i{}^i = 2d/\Ls^2$, and $R_{ij}{}^{ij} = 2/\Ls^2$. Imposing the extremal surface condition $K^i = 0$, we obtain
\begin{equation}\label{eq:HEEQCG}
S_{\text{QG}}(A)= a_d\frac{\text{Area}(\Sigma_A)}{4 G_N} - \frac{\lambda_3}{2 G_N} \int_{\Sigma_A} K^i{}_{ab} K_i{}^{ab}\,,
\end{equation}
where the coefficient $a_d$ is given by $a_d = 1 + 2[d(d+1)\lambda_1 + d\lambda_2 + 2\lambda_3]/\Ls^2$.

The first term in Eq.~\eqref{eq:HEEQCG} coincides with that of the previous section, up to an overall rescaling of the coefficient. The second term, on the other hand, is computationally more involved. Nevertheless, its contribution to the universal part of the pseudoentropy follows the same structural pattern as the area contribution. In particular, the timelike and spacelike sectors exhibit parity-dependent terms that cancel each other. Summing all contributions, we recover the same structure as in Eqs.~\eqref{eq:Sdef1} and \eqref{eq:Sdef2}, with the only differences that $\left.a^\star\right|_{\text{QG}} = a_d \left.a^\star\right|_{\text{E}}$, where the label $\left. \right|_{\text{E}}$ denotes the Einstein-dS result, and, at quadratic order in $\epsilon$, we observe
\begin{equation} \label{eq:CTQCG}
\left.C_T\right|_{\text{QG}} = \left(a_d - \frac{4(d-2)\lambda_3}{\Ls^2}\right)\left.C_T\right|_{\text{E}}\,.
\end{equation}
Similar to Einstein gravity, this result again shows that under $\left.\Ls\right|_{\text{AdS}}\rightarrow-\iu \left.\Ls\right|_{\text{dS}}$, the coefficient $C_T$ for higher-curvature gravity transforms in the same way: $\left.C_T\right|_{\text{QG; AdS}}\rightarrow\left.C_T\right|_{\text{QG; dS}}$.

\noindent\emph{Discussion}. In this Letter, we computed the universal part of the pseudoentropy in dS/CFT for slightly deformed spherical entangling surfaces. We found that the leading correction appears at order $\epsilon^2$---establishing the sphere as a local extremum---and is governed by a coefficient $C_T$ which we conjecture to be the two-point stress-tensor coefficient  of the non-unitary dual CFT. Our findings are consistent with the analytic continuation $\left.\Ls\right|_{\text{AdS}}\rightarrow-\iu \left.\Ls\right|_{\text{dS}}$, supporting the interpretation that part of the holographic data of the non-unitary CFT dual to dS can be obtained via continuation from AdS. These results extend to higher-curvature theories, in particular quadratic curvature gravity, indicating that the structure of the perturbative corrections is universal across non-unitary CFTs with holographic duals.

However, it should be emphasized that this is not a trivial result, as there are examples of operators, such as scalar primaries, whose two-point function coefficients are not obtained by analytic continuation, as explicitly checked in dS$_3$ in \cite{Doi:2024nty}. In fact, all our results and conclusions on the shape dependence of pseudoentropy for small perturbations around the sphere resemble those for holographic entanglement entropy in AdS/CFT.

Recently, in the context of timelike entanglement entropy in AdS, it has been argued that the correct RT surface requires an appropriate complexification of the bulk geometry~\cite{Heller:2024whi}. This viewpoint was further sharpened in dS$_3$/CFT$_2$ by consistency with the first law of pseudoentropy~\cite{Fujiki:2025rtx}, which selects extremal surfaces extending into complexified directions of dS space. This suggests that a fully complexified bulk description is necessary, particularly in the presence of perturbations of the dS metric.

However, a crucial observation, emphasized in Ref.~\cite{Narayan:2026wzp}, is that in pure dS space the area of extremal surfaces depends only on the endpoints, and not on the precise choice of contour in the complex time plane. This leads to a degeneracy among saddles: piecewise timelike+Euclidean contours and fully complexified contours are related by contour deformations and yield identical pseudoentropy. While the timelike+Euclidean contour admits a direct geometric interpretation in dS space, complex contours can be understood as equivalent representations obtained via such deformations. This picture is consistent with the description of no-boundary extremal surfaces and persists as long as one remains in the undeformed background.

It therefore follows that the piecewise extremal surface should be viewed as a representative of this class of deformable saddles. This explains why our construction is expected to reproduce the universal term without explicitly invoking complexified geometries. However, once perturbations of the metric are introduced, this degeneracy is generally lifted, making complex extremal surfaces essential~\cite{Fujiki:2025rtx}. Given the limited understanding of such surfaces in higher-dimensional dS space, we restrict to the piecewise real construction, while our matching of the analytically continued $C_T$ with what was computed for AdS ~\cite{Mezei:2014zla,Faulkner:2015csl} suggests that the equivalence persists in higher dimensions.

We conclude by discussing some future directions. Let us first remark the following successful followups for unitary CFTs: i) the coefficient $C_T$ was shown to control the leading perturbation around the sphere for general unitary CFTs \cite{Faulkner:2015csl}; ii) in $d=3$, the sphere was identified as a global extremum for both holographic \cite{Fonda:2015nma,Anastasiou:2020smm} and general unitary CFTs \cite{Bueno:2021fxb}; iii) also, in $d=3$, the quotient $C_T/F_0$, with $F_0=-\log Z_{\mathbb S^d}$ being the Euclidean free energy when the unitary CFT is placed on the round sphere \cite{Dowker:2010yj,Casini:2011kv}, was conjectured to be bounded by the values of the free boson and the Maxwell theories \cite{Bueno:2023gey,Bueno:2026hku}, and finally; iv) the $\epsilon^4$ correction was found to be characterized by the universal coefficient $t_4$ appearing in the three-point stress-tensor correlator for certain holographic models \cite{Anastasiou:2022pzm}, although this connection does not persist for general theories \cite{Bueno:2015ofa}. Based on these findings, it would be interesting to explore whether they hold for non-unitary CFTs. However, it is important to note that the landscape of non-unitary CFTs is considerably broader than its unitary counterpart. In particular, certain non-unitary theories exhibit $C_T = 0$, implying that our expression \eqref{eq:Sdef2} does not capture their behavior. This rules out a straightforward extension of i) and iii) to general non-unitary CFTs. Consequently, in what follows, we focus on non-unitary CFTs within the same universality class as holographic theories. For this subset, it would be interesting to examine, in line with ii), whether the local extremum of the sphere in $d=3$ remains global, as it occurs in both holographic and general unitary CFTs. Following iii), it would also be worthwhile to verify whether more general higher-curvature gravities exhibit the same universal behavior observed in the quadratic case. Such theories may further provide a framework to study higher $n$-point functions of the stress-energy tensor in the non-unitary dual CFT. This direction was pursued for entanglement entropy in AdS/CFT in \cite{Anastasiou:2022pzm}, where the three-point coefficient $t_4$ was unambiguously determined in cubic curvature gravity.

It would be interesting to extend our analysis to deformed timelike subregions in the dual unitary CFT of AdS, which naturally lead to the notion of timelike entanglement entropy (properly also understood as pseudoentropy)---see e.g., \cite{Doi:2022iyj,Chen:2023gnh,Das:2023yyl,Doi:2023zaf,Guo:2024lrr,Grieninger:2023knz,Heller:2024whi,Jiang:2023ffu,Li:2022tsv,Narayan:2023ebn}. Exploring this construction in AdS and extracting the associated universal holographic data would provide a valuable consistency check against standard holographic entanglement entropy. We expect that the universal quantities---such as the central charge and the coefficient $C_T$---will be reproduced, since they are intrinsic properties of the underlying theory and should not depend on the specific nature of the chosen subregion.

Finally, in this work, we have focused on the universal part of pseudoentropy while omitting divergent, nonuniversal contributions. A refined approach based on the renormalized pseudoentropy would extract the universal term directly. In an upcoming work \cite{Anastasiou:2026bbf}, we will show that the conformal renormalization prescription, previously successful for AdS entanglement entropy \cite{Anastasiou:2022ljq,Anastasiou:2024rxe}, extends naturally to dS, providing a unified definition of the renormalized pseudoentropy.

\noindent\emph{Acknowledgements}. We thank Pablo Bueno, Javier M. Magán, and Tadashi Takayanagi for interesting discussions. This work is partially funded by ANID FONDECYT grants 11240059, 1240043, 1261016, 1231133, and 3230626. I.J.A. gratefully acknowledges support from the Simons Center for Geometry and Physics, Stony Brook University, at which some of the research for this work was performed. A.D. thanks the Departamento de Física y Astronomía at Universidad Andrés Bello and the Departamento de Ciencias at Universidad Adolfo Ibáñez for their hospitality during this work. The work of A.D. is supported by Becas de postgrado UNAP. The work of J.M. is supported by the Beatriu de Pinós fellowship BP 2024 00033 of the Agència de Gestió d'Ajuts Universitaris i de Recerca, Generalitat de Catalunya.

\onecolumngrid
\bibliographystyle{apsrev4-2}
\bibliography{Biblio}
\end{document}